\documentclass[aps,twocolumn,showpacs,superscriptaddress]{revtex4}
\usepackage{graphicx}
\begin{document}
\newcommand{\beq}{\begin{equation}}
\newcommand{\eeq}{\end{equation}}
\newcommand{\ben}{\begin{eqnarray}}
\newcommand{\een}{\end{eqnarray}}
\newcommand{\bea}{\begin{array}}
\newcommand{\eea}{\end{array}}
\newcommand{\om}{(\omega )}
\newcommand{\bef}{\begin{figure}}
\newcommand{\eef}{\end{figure}}
\newcommand{\leg}[1]{\caption{\protect\rm{\protect\footnotesize{#1}}}}
\newcommand{\ew}[1]{\langle{#1}\rangle}
\newcommand{\be}[1]{\mid\!{#1}\!\mid}
\newcommand{\no}{\nonumber}
\newcommand{\etal}{{\em et~al }}
\newcommand{\geff}{g_{\mbox{\it{\scriptsize{eff}}}}}
\newcommand{\da}[1]{{#1}^\dagger}
\newcommand{\cf}{{\it cf.\/}\ }
\newcommand{\ie}{{\it i.e.\/}\ }

\title{ Open system of interacting fermions:\\
Statistical properties of cross sections and fluctuations}

\author{G.L.~Celardo}
\affiliation{Instituto de F\'{\i}sica, Universidad Aut\'{o}noma de
Puebla, Apartado Postal J-48, Puebla, Pue., 72570, M\'{e}xico}
\author{F.M.~Izrailev}
\affiliation{Instituto de F\'{\i}sica, Universidad Aut\'{o}noma de
Puebla, Apartado Postal J-48, Puebla, Pue., 72570, M\'{e}xico}
\author{V.G.~Zelevinsky}
\affiliation{
NSCL and Department of Physics and Astronomy, Michigan State
University, East Lansing, Michigan 48824-1321, USA.
}
\author{G.P.~Berman}
\affiliation{
Theoretical Division and CNLS, Los Alamos National
Laboratory, Los Alamos, New Mexico 87545, USA.
}

\begin{abstract}
Statistical properties of cross sections are studied for an open
system of interacting fermions. The description is based on the
effective non-Hermitian Hamiltonian that accounts for the
existence of open decay channels preserving the unitarity of the
scattering matrix. The intrinsic interaction is modelled by the
two-body random ensemble of variable strength. In particular, the
crossover region from isolated to overlapping resonances
accompanied by the effect of the width redistribution creating
super-radiant and trapped states is studied in detail. The
important observables, such as average cross section, its
fluctuations, autocorrelation functions of the cross section and
scattering matrix, are very sensitive to the coupling of the
intrinsic states to the continuum around the crossover. A detailed
comparison is made of our results with standard predictions of
statistical theory of cross sections, such as the Hauser-Feshbach
formula for the average cross section and Ericson theory of
fluctuations and correlations of cross sections. Strong deviations
are found in the crossover region, along with the dependence on
intrinsic interactions and degree of chaos inside the system.
\end{abstract}

\date{\today}
\maketitle

\section{Introduction}

Information on properties of quantum mesoscopic systems comes mostly
from various reactions where the system plays a role of a target. At
high density of intrinsic states, the dynamics of realistic systems
of interacting constituents becomes chaotic. Onset of chaos
immensely complicates the details of the scattering process as
reflected in the paradigm of compound nucleus. At low energies, the
long-lived resonance states are exceedingly complex superpositions
in the basis of independent particles. As energy increases, the
scattering pattern evolves from the set of narrow isolated
resonances to overlapping resonances and strongly fluctuating cross
sections. Since the individual properties of resonances cannot be
predicted, only statistical description is practical and sensible.
The average cross sections are usually described according to
Hauser-Feshbach \cite{hauser52}, while the fluctuations and
correlations of cross sections are treated in terms of Ericson
theory \cite{ericson63,brink63,EMK66}.

Standard theory of statistical reactions does not answer the
question of interplay between reactions and internal structure
determined by the character of interactions between the
constituents. Here more detailed considerations are required based
on the generalization of the shell model of nuclear reactions
\cite{MW}. Such an extension introduces statistical assumptions
concerning intrinsic dynamics and its coupling to the continuum
\cite{agassi75,verbaarschot85} with the extreme limiting case of
intrinsic chaos modelled by the Gaussian Orthogonal Ensemble (GOE)
\cite{brody81}. An important element is still missing here, namely
the transitional regime from separated to overlapping resonances.
The consistent description based on the continuum shell model
\cite{MW}, as well as more phenomenological approaches
\cite{moldauer67,moldauer69,simonius74}, indicate the presence of
a sharp restructuring of the system when the widths of resonances
become comparable to their energy spacings. This phenomenon
carries features of a quantum phase transition with the strength
of continuum coupling playing the role of a control parameter.

As was clearly observed in the shell model framework
\cite{kleinwachter85}, the distribution of resonance widths rapidly
changes in the transitional region in such a way that a number of
very broad resonances equal to a number of open decay channels
absorb the lion's share of the total width of all overlapped
resonances, while the remaining states become very narrow. The
corresponding theory was suggested in \cite{SZPL88,SZNPA89,SZAP92},
where the mechanism of this restructuring was understood to be
associated with the nature of the effective non-Hermitian
Hamiltonian \cite{MW} that describes the intrinsic dynamics after
eliminating the channel variables. The factorized structure of this
Hamiltonian, in turn, is dictated by the unitarity of the scattering
matrix \cite{durand76}. One can compare this phenomenon to the
classical factorized model \cite{BBolsterli59} of a giant resonance,
where the collective strength of many particle-hole states is
shifted in energy and concentrated at a specific combination of
excited states. In spite of formal analogy, physics under study here
is different. The concentration of widths on a few broad states can
be described as collectivization along the {\sl imaginary axis} in
the complex energy plane. The driving force of this restructuring is
the presence of open decay channels and interaction of intrinsic
states through the continuum. The intrinsic Hermitian interaction is
present as well and should be fully accounted for; one of the goals
of our study is to understand the dependence of the continuum
picture on the strength and character of interactions inside the
closed system. The interplay of two collectivities is an interesting
subject \cite{SZAlaga90,sokolov97} practically important in relation
to the so-called pygmy-resonances in loosely bound systems
\cite{varenna06}.

The segregation of short-lived broad resonances from long-lived
trapped states was shown to be similar to the superradiance
\cite{dicke54} in quantum optics induced by the coupling of atomic
radiators through the common radiation field, an analog of coherent
coupling of many overlapped intrinsic states through continuum.
Later a general character of the phenomenon was demonstrated for
systems with GOE intrinsic dynamics and many open channels
\cite{ISS94,SFT99}. Modern versions of the shell model in continuum
\cite{rotter91,VZCSM05} are based on the effective Hamiltonian and
naturally reveal the superradiance phenomenon as an important
element. The transition to this regime should be taken into account
in all cases when a physical system is strongly coupled to the
continuum, see for example \cite{VZWNMP04} and references therein.

The segregation of scales is spectacularly seen in level and width
statistics \cite{SZNPA89,ISS94,haake,mizutori93}. Even for GOE
intrinsic dynamics, the probability $P(s\rightarrow 0)$ of very
small spacings between the centroids of resonances does not vanish
because of energy uncertainty of unstable states. The width
distribution reveals the separation of the ``cloud" of superradiant
states far from the real energy axis, while the trapped states are
clearly accumulated near the real axis. Some features of the picture
are however sensitive to the character and strength of intrinsic
interactions as we have demonstrated recently \cite{celardo1}. The
analysis included the two-body random ensemble of variable
interaction strength in a Fermi system of shell-model type and the
GOE as an extreme limit of many-body random interaction.

In the present work, using the same framework as in \cite{celardo1},
we study the interplay between the intrinsic dynamics and
statistical properties of cross sections comparing the results with
those of conventional approaches, namely Hauser-Feshbach average
cross sections and Ericson fluctuations and correlations. In
particular, we show that the assumption that fluctuations of the
resonance widths are negligible for a large number of channels is
not correct. We also show that the elastic enhancement factor
strongly depends on the degree of chaoticity inside the system, thus
leading to deviations from the Hauser-Feshbach formula.

\section{The model}

\subsection{Hamiltonian}

We consider a system of $n$ interacting fermions on $m$ mean-field
orbitals. A large number, $N=m!/[n!(m-n)!]$, of intrinsic many-body
states $|i\rangle$ comprise our Hilbert space. In our simulations we
take $n=6,\,m=12$ that provides a sufficiently large dimension
$N=924$. The states are unstable being coupled to $M$ open decay
channels. The dynamics of the whole system is governed by an
effective non-Hermitian Hamiltonian \cite{MW,SZNPA89,VZWNMP04} given
by a sum of two $N\times N$ matrices,
\begin{equation}
{\cal H}= H - \frac{i}{2}\, W\,;\,\,\,\,\,\,\,\,\,\,\, W_{ij}=\sum
_{c=1}^M A_{i}^{c}A_j^c.                         \label{1}
\end{equation}
Here and below the intrinsic many-body states are labelled as
$i,j,...$ and decay channels as $a,b,c...$.  In Eq. (\ref{1}), $H$
describes Hermitian internal dynamics that in reality also can be
influenced by the presence of the continuum
\cite{rotter91,VZCSM05,CSM06}, while $W$ is a sum of terms
factorized in amplitudes $A^c_i$ coupling intrinsic states
$|i\rangle$ to the channels $c$. Under time-reversal invariance,
these amplitudes can be taken as real quantities so that both $H$
and $W$ are real symmetric matrices.

We model $H$ by the two-body random ensemble (TBRE) assuming the
intrinsic Hamiltonian $H$ in the form $H=H_0+V$, where $H_0$
describes the mean-field single-particle levels $|\nu \rangle$,
and $V$ is a random two-body interaction between the particles
\cite{FI97}. The {\sl single-particle} energies, $\epsilon_\nu$,
are assumed to have a Poissonian distribution of spacings, with
the mean level density $1/d_0$. The interaction $V$ is
characterized by the variance of the {\sl two-body} random matrix
elements, $\langle V_{\nu_1,\nu_2;\nu_3,\nu_4}^2\rangle=v_0^2$.
With no interaction, $v_0=0$, the many-body states have also the
Poissonian spacing distribution $P(s)$. In the opposite extreme
limit, $d_0=0$, corresponding to infinitely strong interaction,
$\lambda\equiv v_0/d_{0} \rightarrow \infty$, the function $P(s)$
is close to the Wigner-Dyson (WD) distribution typical for a
chaotic system \cite{brody81}. Following Ref. \cite{FI97}, the
critical interaction for the onset of strong chaos can be
estimated as
\begin{equation}
\label{estimate} \lambda_{{\rm cr}}=\frac{v_{{\rm cr}}}{d_0} \approx
\frac{2(m-n)}{N_s},
\end{equation}
where $N_s=n(m-n)+ n(n-1)(m-n)(m-n-1)/4$ is the number of directly
coupled many-body states in any row of the matrix $H_{ij}$. Thus,
we have $\lambda_{{\rm cr}}\approx 1/20$, and often we perform the
simulations with the value $\lambda=1/30$ slightly lower than
$\lambda_{{\rm cr}}$. In parallel, we also consider the intrinsic
Hamiltonian $H$ belonging to the GOE that corresponds to a {\sl
many-body} interaction, when the matrix elements are Gaussian
random variables, $\langle H_{ij}^2\rangle=1/N$ for $i \ne j$ and
$\langle H_{ij}^2\rangle=2/N$ for $i= j$.

The real amplitudes $A_i^c$ are assumed to be random independent
Gaussian variables with zero mean and variance
\begin{equation}
\langle A_i^c A^{c'}_j\rangle=\delta_{ij}\delta^{cc'}\,\frac{
\gamma^{c}}{N}.                                \label{2}
\end{equation}
The parameters $\gamma^{c}$ with dimension of energy characterize
the total coupling of all states to the channel $c$. The
normalization used in Eq. (\ref{2}) is convenient if the energy
interval $ND$ covered by decaying states is finite. Here $D$ is the
distance between the many-body states in the middle of the spectrum,
$D=1/\rho(0)$, where $\rho(E)$ is the level density, and $E=0$
corresponds to the center of the spectrum. We neglect a possible
explicit energy dependence of the amplitudes that is important near
thresholds and is taken into account in realistic shell model
calculations \cite{VZCSM05,CSM06}. The ratio $\gamma^{c}/ND$
characterizes the degree of overlap of the resonances in the channel
$c$. We define the corresponding control parameter as
\begin{equation}
\kappa^{c}=\frac{\pi\gamma^{c}}{2ND}.              \label{3}
\end{equation}
The transitional region corresponds to $\kappa^{c}\approx 1$.
Varying the intrinsic interaction and, therefore, the level density
$\rho$, we renormalize correspondingly the absolute magnitude of the
widths, $\gamma$, in order to keep the coupling to continuum given
by Eq. (\ref{3}) fixed.

\subsection{Scattering matrix}

The effective Hamiltonian allows one to study the cross sections for
possible reactions $b\rightarrow a$,
\begin{equation}
\sigma^{ba}(E) = |{\cal T}^{ba}(E)|^2.            \label{4}
\end{equation}
(our cross sections are dimensionless since we omit the common
factor $\pi/k^{2}$). In what follows we study both the elastic,
$b=a$, and non-elastic, $b \neq a$, cross sections. Ignoring the
smooth potential phases irrelevant for our purposes we express the
scattering amplitude of the reaction, ${\cal T}^{ba}$, in terms of
the amplitudes $A_{i}^{c}$,
\begin{equation}
{\cal T}^{ba}(E)=\sum_{i,j}^N A_i^b\left(\frac{1}{E-{\cal H}}
\right)_{ij} A_j^a.                          \label{5}
\end{equation}
Here the denominator contains the total effective Hamiltonian
(\ref{1}) including in this way the continuum coupling $W$ to all
orders.

We can also write ${\cal T}^{ba}(E)$ in a different way,
diagonalizing the effective non-Hermitian Hamiltonian ${\cal H}$.
Its eigenfunctions $|r\rangle$ and $\langle \tilde{r}|$ form a
bi-orthogonal complete set,
\begin{equation}
{\cal H}|r\rangle={\cal E}_{r}|r\rangle, \quad \langle \tilde{r}|
{\cal H}=\langle\tilde{r}|{\cal E}^{\ast}_{r},     \label{6}
\end{equation}
and its eigenvalues are complex energies,
\begin{equation}
{\cal E}_{r}= E_{r}-\,\frac{i}{2}\,\Gamma_{r},    \label{7}
\end{equation}
corresponding to the resonances with centroids $E_{r}$ and widths
$\Gamma_{r}$. The decay amplitudes $A^{b}_{i}$ are transformed
according to
\begin{equation}
{\cal A}^{b}_{r}=\sum_{i}A^{b}_{i}\langle i|r\rangle, \quad
\tilde{{\cal A}}^{a}_{r}=\sum_{j}\langle \tilde{r}|j\rangle
A^{a}_{j},                                        \label{8}
\end{equation}
and the transition amplitudes are given by
\begin{equation}
{\cal T}^{ba}(E)= \sum_{r}^N {\cal A}^{b}_{r}\, \frac{1}{E-{\cal
E}_r}\tilde{{\cal A}}^{a}_{r},                 \label{9}
\end{equation}
The bi-orthogonality of the transformation ensures that the
statistical properties (\ref{2}) of the ensemble of the amplitudes
are preserved,
\begin{equation}
\langle\tilde{{\cal A}}^{a}_{r}{\cal A}^{b}_{r'}\rangle=\delta^{ab}
\delta_{rr'}\,\frac{\gamma^{a}}{N}.            \label{10}
\end{equation}

Introducing the matrix in channel space analogous to what is
routinely used in the resonance data analysis,
\begin{equation}
K=\frac{1}{2}\, {\bf A}\, \frac{1}{E-H}\,{\bf A}^{T}, \label{11}
\end{equation}
where the denominator includes only intrinsic dynamics and ${\bf A}$
is the $N \times M$ matrix of the transition amplitudes $A^{c}_{j}$,
one can relate it to the transition amplitude ${\cal T}$ and the
scattering matrix,
\begin{equation}
S^{ba}=\delta^{ba}-i{\cal T}^{ba},                 \label{12}
\end{equation}
in the explicitly unitary form
\begin{equation}
S=\frac{1-iK}{1+iK}.                            \label{13}
\end{equation}
The poles of the $K$-matrix are real eigenvalues $E_{\alpha}$ of the
intrinsic Hermitian Hamiltonian $H$. The complex eigenvalues
(\ref{7}) coincide with the poles of the $S$-matrix and, for small
values of $\gamma$, determine energies and widths of separated
resonances. With an increase of $\gamma$, the resonances start to
overlap leading to specific features of the scattering process which
are of our main interest.

\section{Average scattering matrix}

The scattering matrix (\ref{13}) averaged over the ensemble of decay
amplitudes (\ref{2})  with accuracy of $1/N$, is given by
\cite{SZNPA89}
\begin{equation}
\langle S\rangle=\frac{1-i\langle K\rangle}{1+i\langle K\rangle}.
                                              \label{14}
\end{equation}
Following Refs. \cite{SZNPA89,gorin02}, we assume, in concordance
with the statistical ansatz (\ref{2}), that the decay amplitudes and
the eigenvalues $E_{\alpha}$ of the intrinsic Hamiltonian are
statistically independent. This assumption is satisfied for the GOE
and Poissonian ensembles, and here we assume that this is also true
for the TBRE.

Since the mean field many-body basis $|i\rangle$ and the
eigenbasis $|\alpha\rangle$ of the intrinsic Hermitian Hamiltonian
are related by a real orthogonal transformation, the
anti-Hermitian part $W$ of the effective Hamiltonian in the basis
$|\alpha\rangle$ still has a factorized form in terms of the new
amplitudes $B^{c}_{\alpha}$; similarly to Eq. (\ref{10}), the
correlation function of the amplitudes $B$ coincides with that
given by Eq.~(\ref{2}). Thus, the average $K$-matrix determined by
the statistical properties of the decay amplitudes reduces to
\begin{equation}
\langle K^{ab}(E)\rangle=\frac{1}{2}\,\sum_{\alpha}\frac{\langle
B^{a}_{\alpha} B^{b}_{\alpha}\rangle}{E-E_{\alpha}} =\delta^{ab}
\,\frac{\gamma^{a}}{2 N}\,\sum_{\alpha}\frac{1}{E-E_{\alpha}}.
                                             \label{15}
\end{equation}

The last sum in Eq.(\ref{15}) is the trace of the intrinsic Green
function $1/(E-H)$. For energy $E$ inside the spectrum of $H$ we
should understand it as a limiting value, $E\rightarrow E+i0$. The
trace of the imaginary part of the Green function determines the
level density $\rho(E)=\sum_{\alpha} \delta(E-E_{\alpha})$ for the
Hamiltonian $H$, and
\begin{equation}
\sum_{\alpha}\frac{1}{E-E_{\alpha}+i0}={\rm P.v.}\,\sum_{\alpha}
\frac{1}{E-E_{\alpha}}-i\pi\rho(E).            \label{16}
\end{equation}
The principal value part is a smooth function of energy that
vanishes in the middle of the spectrum; as a result, in this
vicinity
\begin{equation}
\langle K^{ab}\rangle=-i\pi\delta^{ab}\,\frac{\gamma^{a}}{2N}\,
\rho(0)=-i\delta^{ab}\kappa^{a},               \label{17}
\end{equation}
where $\kappa^{a}$ is defined by Eq. (\ref{3}).

The  average  scattering matrix (\ref{14}) takes the form
\begin{equation}
\langle S^{ab}\rangle=\delta^{ab}\,\frac{1-\kappa^{a}}
{1+\kappa^{a}}                                    \label{18}
\end{equation}
that depends only on our parameter of continuum coupling
independently of the intrinsic interaction strength $\lambda$. This
independence is due to the fact that we normalized $\kappa$ to the
mean level spacing, $D$, but it is important to stress that in the
shell model $D$ does depend on the intrinsic interaction \cite{big}.
Eq.(\ref{18}) implies that the transmission coefficient,
\begin{equation}
T^{a}=1-|\langle S^{aa}\rangle|^2,               \label{19}
\end{equation}
is also independent of this strength,
\begin{equation}
T^{a}=\frac{4\kappa^{a}}{(1+\kappa^{a})^2}.               \label{20}
\end{equation}
The transmission coefficient in the channel $a$, $T^{a}$, is maximum
(equal to 1) at the critical point of this channel, $\kappa^{a}=1$,
when the average $S$-matrix vanishes. Thus, $\kappa=1$ determines
the so-called {\sl perfect coupling regime}. We will study the
statistical properties of cross sections as a function of the
intrinsic interaction strength $\lambda$ and continuum coupling
parameter $\kappa$, both below the critical point, $\kappa<1,$ and
after the superradiance transition has occurred, $\kappa>1$.

\section{Comparing the ensembles}

The  density of states of the GOE ensemble follows the famous
semicircle law, while the density of states of the TBRE is
Gaussian for large enough particle number, $n$, and orbital
number, $m$, \cite{brody81}; its width depends on $\lambda$. In
order to compare different ensembles, we restrict our statistical
analysis to a small energy interval with a constant level density
at the center of the real spectrum of the complex eigenvalues of
${\cal H}$. This interval should be small enough in order to
neglect the energy-dependent difference of the density of states
among the ensembles, but large enough with respect to the widths
in order to contain a statistically meaningful number of
resonances.

For a model with a finite resonance number, it is important to avoid
edge effects, see discussion in \cite{moldauer67,moldauer75}. The
energy interval subject to statistical analysis should be also at a
distance of at least several widths away from the edges. A rough
estimate \cite{celardo1} goes as follows: for $M$ equivalent
channels, $\Gamma/D \propto M$, and the distance from the center to
the edges is $ND/2$, then $ND/2\gg\Gamma/D$ that implies $M/N \ll
1/2$. This shows that the ratio of the number of channels to that of
resonances must be small in order for the results to be model
independent. With this choice, the model will be essentially
equivalent to an infinite resonance model with a constant level
density, apart from a narrow interval around the critical value. For
$\kappa=1$, with an infinite resonance number, the average widths
should logarithmically diverge, in agreement with the
Moldauer-Simonius expression \cite{simonius74}. In our finite model,
the results become model-dependent in a narrow interval near
$\kappa=1$. Our approach is still appropriate for a comparison with
the predictions of Ericson fluctuation theory derived for an
infinite resonance model with a constant level density.

The results of numerical simulations presented below refer to the
case of $N=924$ internal states and $M$ equiprobable channels,
$\kappa^{a}=\kappa$. The maximum value of $M$ we considered is
$M=25$ so that $M/N \approx 2\cdot 10^{-2}$. For any value of
$\kappa$, we have used a large number of realizations of the
Hamiltonian matrices, with further averaging over energy.

\section{Ericson fluctuations}

The starting point of the conventional theory
\cite{ericson63,brink63,EMK66} can be summarized as follows. The
scattering amplitude ${\cal T}^{ab}(E)=\langle{\cal
T}^{ab}(E)\rangle + {\cal T}^{ab}_{{\rm fl}}(E)$ is divided into
two parts, an average one, $\langle{\cal T}^{ab}(E)\rangle$, and a
fluctuating one, ${\cal T}^{ab}_{{\rm fl}}(E)$, with
\begin{equation}
\langle{\cal T}^{ab}_{{\rm fl}}(E)\rangle =0. \label{21}
\end{equation}
Note that in our statistical model we have $\langle {\cal T}_{{\rm
inel}} \rangle=0$ for inelastic channels, while
\begin{equation}
\langle {\cal T}_{{\rm el}} \rangle=-i(1-\langle S
\rangle)=-2i\,\frac{\kappa} {1+\kappa}            \label{39}
\end{equation}
for elastic channels.

With statistical independence of poles (resonance energies) and
residues (resonance amplitudes), $z_r\equiv{\cal A}_r^a \tilde{{\cal
A}}_r^b$, we obtain $z_r=\langle z_r \rangle+ \delta z_r$, so that
\begin{equation}
{\cal T}^{ab}(E)= \sum_r \frac{\langle z_r \rangle+\delta
z_r}{E-E_r+i \Gamma_r/2}.                     \label{22}
\end{equation}
In the regime of overlapping resonances, $\langle\Gamma\rangle > D$,
and assuming all widths of the same order, $\Gamma_r \sim \langle
\Gamma \rangle$, the average part can be computed similarly to Eq.
(\ref{16}), substituting the sum by the integral,
\begin{equation}
\int\frac{\rho(E_r)\langle z_r\rangle dE_r}{E-E_r+i\langle \Gamma
\rangle /2} \approx -i \frac{\pi \langle z_r \rangle}{D},
\label{23}
\end{equation}
where a constant level density, $\rho(E_r)=1/D$, is assumed.

The average cross section, $\sigma= |{\cal T}|^2$, also can be
divided into two contributions,
\begin{equation}
\langle\sigma\rangle=\langle|{\cal T}|^2\rangle=|\langle {\cal T}
\rangle|^2+\langle|{\cal T}_{{\rm fl}}|^2\rangle.  \label{24}
\end{equation}
The two terms in Eq. (\ref{24}) are interpreted as
\begin{equation}
\langle\sigma\rangle=\langle\sigma_{{\rm dir}}\rangle + \langle
\sigma_{{\rm fl}}\rangle,                        \label{25}
\end{equation}
where the direct reaction cross section, $\langle \sigma_{{\rm dir}}
\rangle$, is determined by the average scattering amplitude only,
while $\langle \sigma_{{\rm fl}} \rangle$ is the fluctuational cross
section (also called in the literature the compound nucleus cross
section) that is determined by the fluctuational scattering matrix.

For overlapping resonances, $\langle \Gamma \rangle > D$, the
following conclusions were derived concerning the scattering
amplitude and the statistical properties of the cross sections.

(A). {\sl The average fluctuational cross section} \cite{ericson63}.
Assuming that $\Gamma_r \approx \langle\Gamma\rangle$ for a large
number of channels, i.e. fluctuations of the widths around their
average value are small, ${\rm Var}(\Gamma)/\langle\Gamma \rangle^2
\ll 1$, the average fluctuational cross section,
$\langle\sigma_{{\rm fl}} \rangle=\langle{\cal T}_{{\rm fl}}{\cal
T}^*_{{\rm fl}}\rangle$, can be written as
\begin{equation}
\langle\sigma_{{\rm fl}}\rangle=\left\langle\sum_{rr'}\frac{\delta
z_r^{\ast} \delta z_{r'}}{(E-E_r+i/2\langle\Gamma\rangle)(E-E_r'-i/2
\langle \Gamma \rangle)}\right \rangle,                  \label{26}
\end{equation}
where the substitution $\Gamma_r\approx\langle\Gamma \rangle$ was
used. Now the averaging over energy is applied,
\begin{equation}
\langle F(E) \rangle \Rightarrow \frac{1}{\Delta E}\, \int dE\,
F(E).                                             \label{27}
\end{equation}
The integration leads to
\begin{equation}
\langle\sigma_{{\rm fl}}\rangle= \frac{2\pi i}{\Delta E}\sum_{rr'}
\frac{\delta z_r^{\ast} \delta z_{r'}}{E_{r'}-E_r+i\langle\Gamma
\rangle}.
                                                 \label{28}
\end{equation}
Now we assume that $\delta z_{r}$ are uncorrelated random quantities
with the statistics independent of $r$, $\langle \delta
z_{r}^{\ast}\delta z_{r'}\rangle=\delta_{rr'}\langle |\delta
z|^{2}\rangle$. The absence of correlations between the amplitudes
$\delta z_r$ gives
\begin{equation}
\langle\sigma_{{\rm fl}} \rangle = \frac{2 \pi}{D} \frac{\langle
|\delta z|^2 \rangle}{\langle \Gamma \rangle}, \quad D=\frac{\Delta
E}{N}.                                \label{29}
\end{equation}

(B). {\sl Variance of the cross section}, ${\rm Var}(\sigma)=\langle
\sigma^2 \rangle- \langle \sigma \rangle^2$. The derivation can be
performed under more general assumptions \cite{ericson60} than those
used for the analysis of average cross section. If the quantities
$\delta z_r /(E- {\cal E}_r)$ are independent complex variables,
then ${\cal T}$ is Gaussian distributed, that is ${\cal T}= \xi +i
\eta $, where both $\xi$ and $\eta$ are Gaussian random variables
with zero mean. This is due to the fact that for $\langle \Gamma
\rangle \gg D$ both $\xi$ and $\eta$ are the sums of a large number
of random variables. In the conventional theory it is also assumed
that $\xi$ and $\eta$ have equal variance.

Then for the fluctuating cross section we have
\begin{equation}
\langle\sigma_{{\rm fl}}^2 \rangle= \langle |{\cal T}_{{\rm fl}}|^4
\rangle= \langle {\cal T}_{{\rm fl}}{\cal T}_{{\rm fl}}^*{\cal
T}_{{\rm fl}}{\cal T}_{{\rm fl}}^*\rangle= 2\langle\sigma_{{\rm fl}}
\rangle^2.                                         \label{30}
\end{equation}
In a more general case when $\langle {\cal T} \rangle \ne 0$,
\begin{equation}
{\rm Var}(\sigma)=\langle\sigma_{{\rm fl}}\rangle \left( 2 \langle
\sigma_{{\rm dir}}\rangle+\langle\sigma_{{\rm fl}}\rangle \right).
                                              \label{31}
\end{equation}

(C). {\sl The correlation function of the scattering amplitudes} is
defined as
\begin{equation}
c(\epsilon)=\langle{\cal T}(E+\epsilon){\cal T}^*(E)\rangle -
|\langle{\cal T}(E)\rangle|^2=\langle{\cal T}_{{\rm fl}}(E+\epsilon)
{\cal T}_{{\rm fl}}^*(E) \rangle.              \label{32}
\end{equation}
Evaluating $c(\epsilon)$ under the same assumptions as for the
average cross sections, one obtains,
\begin{equation}
c(\epsilon)=\langle\sigma_{{\rm fl}}\rangle\,\frac{\langle \Gamma
\rangle}{\epsilon+i \langle \Gamma \rangle}.     \label{33}
\end{equation}

(D). {\sl The cross section correlation function} is defined as
\begin{equation}
C(\epsilon)=\langle \sigma(E)\sigma(E+\epsilon)\rangle- \langle
\sigma(E) \rangle^2.                         \label{34}
\end{equation}
Taking into account the Gaussian form of distribution for $\cal T$
and Eq.(\ref {33}), one obtains that:

(a) the normalized autocorrelation function of cross sections
satisfies the relation,
\begin{equation}
\frac{C(\epsilon)}{C(0)}=\frac{|c(\epsilon)|^2}{|c(0)|^2};
                                            \label{35}
\end{equation}

(b) the correlation function has a Lorentzian form,
\begin{equation}
\frac{C(\epsilon)}{C(0)}=\frac{l^2}{l^2+\epsilon^2}, \label{36}
\end{equation}
where the correlation length, $l$, is equal to the average width,
\begin{equation}
l=\langle\Gamma\rangle.                          \label{37}
\end{equation}

In the following we compare the predictions (B)-(D) of the
conventional theory of Ericson fluctuations with our numerical
results, paying special attention to the dependence on the
intrinsic interaction strength, $\lambda$. As for the part (A),
since there are no predictions for the quantity $|\delta z|^2$
determining the average fluctuational cross section (\ref{29}),
the comparison will be done with the Hauser-Feshbach, theory
widely used in the literature.

\section{Average cross section}

In this section we study how total and partial cross sections
depend on the continuum coupling, $\kappa$, and intrinsic
interaction, $\lambda$. For any value of $\kappa$ we have used
$N_r=30$ realizations of the Hamiltonian matrices. For each
realization we took into account only the interval [-0.2,0.2] of
real energy at the center of the spectrum.

It follows from Eq. (\ref{13}) that the average total cross section
defined by the optical theorem,
\begin{equation}
\langle\sigma_{{\rm tot}}\rangle=2 (1-{\rm Re}\,\langle
S\rangle)=\frac{4 \kappa}{1+\kappa},             \label{40}
\end{equation}
depends only on the average scattering matrix and therefore is
independent of $\lambda$ and $M$. Since $\sigma_{{\rm el}}=
\sigma_{{\rm tot}}$ for $M=1$, the average elastic cross section
is also independent of $\lambda$ for the case of one channel. The
situation changes as we increase the number of channels.

In order to analyze the average elastic and inelastic cross section,
we single out the average scattering matrix elements in the standard
form, $S^{ab}=\langle S^{ab}\rangle+S_{{\rm fl}}^{ab}$, where
$\langle S^{ab}\rangle=\delta^{ab}\langle S^{aa}\rangle$ and
$\langle S_{{\rm fl}}^{ab}\rangle=0$. The average inelastic cross
section, $a\neq b$, is given by
\begin{equation}
\langle \sigma^{{\rm ab}}\rangle =\langle|S^{ab}|^2\rangle=\langle
|S_{{\rm fl}}^{ab}|^2\rangle.                    \label{41}
\end{equation}
The average elastic cross section can be written as
\begin{equation}
\langle \sigma^{aa} \rangle =  |1-\langle
S^{aa}\rangle|^2+\langle|S_{{\rm fl}}^{aa}|^2\rangle. \label{42}
\end{equation}
Following the literature we will call $\langle|S_{{\rm
fl}}^{ab}|^2\rangle$ the fluctuational cross section,
$\langle\sigma_{{\rm fl}}^{ab}\rangle$. For $M$ equivalent
channels the fluctuational cross section can be expressed with the
use of the {\sl elastic enhancement factor},
\begin{equation}
F=\frac{\langle \sigma_{{\rm fl}}^{aa}\rangle}{\langle
\sigma_{{\rm fl}}^{ab} \rangle},
\label{43}
\end{equation}
where $b\neq a$. Indeed, in the case of equal channels, using the
relation,
\begin{equation}
\sigma_{{\rm tot}}= \sum_b \sigma^{ab} = (M-1) \sigma^{{\rm
inel}}+|1-S^{aa}|^2,
                                                 \label{44}
\end{equation}
we obtain
\begin{equation}
\langle \sigma_{{\rm fl}}^{ab} \rangle=\frac{1-|\langle
S^{aa}\rangle|^2}{F+M-1}=\frac{T}{F+M-1},       \label{45}
\end{equation}
where $T$ is the transmission coefficient defined in Eq. (\ref{20}),
and
\begin{equation}
\langle \sigma_{{\rm fl}}^{aa} \rangle=F\langle \sigma_{{\rm
fl}}^{ab} \rangle=\frac{FT}{F+M-1};\,\,\,\,\,\,\,\,\,\, a \neq  b
. \label{46}
\end{equation}

Since the transmission coefficient $T$ does not depend on
$\lambda$, the only dependence on $\lambda$ in Eqs.~(\ref{45}) and
(\ref{46}) is contained in the elastic enhancement factor $F$. The
same seems to be correct even when the channels are
non-equivalent, according to the results of Ref.~\cite{muller87}.
Leaving a detailed analysis of the elastic enhancement factor for
a separate study, here we point out that $F$ also depends on
$\kappa$. Specifically, with an increase of $\kappa$ from zero,
the value of $F$ decreases, being confined by the interval between
$3$ and $2$.

\begin{figure}[h!]
\includegraphics[width=8cm,angle=-90]{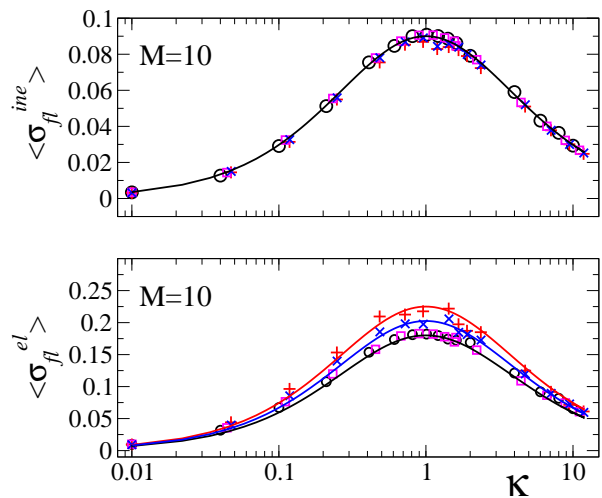}
\vspace{-0.5cm} \caption{(Color online) Fluctuational inelastic
(upper panel) and fluctuational elastic (lower panel) cross sections
as a function of $\kappa$. Circles refer to the GOE case, pluses to
$\lambda=0$, crosses to $\lambda=1/30$, and squares to
$\lambda\rightarrow\infty$. The solid curves (top-down) correspond
to the HF formula (\ref{49}) with $F=2.5,\,2.25,\,2.0$ for
$\lambda=0,\,1/30$ and the GOE, respectively. The values $2.5$ and
$2.25$ are found numerically using the definition (\ref{43}).}
\label{ERAVE}
\end{figure}

For the fluctuational inelastic cross section, with an increase of
the number of channels the dependence on the interaction strength
$\lambda$ disappears, see Fig.~\ref{ERAVE} (upper panel). This is in
agreement with the fact that in the limit of large $M$ we have $
\langle \sigma_{{\rm fl}}^{ab} \rangle \rightarrow T/M$,
independently of $\lambda$, see Eq.~(\ref{45}). In contrast, the
fluctuational elastic cross section manifests a clear dependence on
$\lambda$, see Fig.~\ref{ERAVE} (lower panel). Thus, one can
directly relate the value of the enhancement factor $F$ to the
strength $\lambda$ of interaction between the particles. As one can
see, the more regular is the intrinsic motion, the higher is the
average cross section. Note that for a large number of channels the
$\lambda-$dependence of the elastic cross section is in agreement
with the estimate, $\langle \sigma_{{\rm fl}}^{aa} \rangle
\rightarrow FT/M$, see Eq.~(\ref{46}).

Now we compare our numerical results with the Hauser-Feshbach (HF)
formula derived under quite general assumptions for both $a=b$ and
$a \neq b$ (see Ref. \cite{brody81} and references therein),
\begin{equation}
\langle \sigma_{{\rm fl}}^{ab} \rangle = (1+ \delta_{ab})\,
\frac{T^a T^b}{\sum T^c}= (1+ \delta_{ab})\,\frac{T}{M}.
                                                \label{47}
\end{equation}
Here the last expression corresponds to our case of equivalent
channels, $T^c=T$. As one can see, the HF formula predicts $F=2$,
independently of the interaction strength between the particles.
This formula was also derived in \cite{agassi75} in the
overlapping regime for the TBRE ensemble with the infinite
interaction, and in \cite{lehmann95} for the GOE ensemble. For
finite number of channels in the limit $1/M \ll 1$, the corrected
HF formula was derived in \cite{agassi75}. For equivalent channels
it reads,
\begin{equation}
\langle\sigma_{{\rm fl}}^{ab}\rangle= (1+ \delta_{ab})\,\frac{T}{M}
\left( 1-\frac{1}{M}\right).                        \label{48}
\end{equation}

Our data confirm that for the fluctuational {\sl inelastic} cross
section the HF formula gives correct results for all values of
$\lambda$ in the case of large number of channels. The specific case
of small number of channels, for which the HF is not valid, will be
discussed elsewhere.

On the other hand, for the fluctuational {\sl elastic} cross
section, our data show that the HF formula works only in the GOE
case and in the limit $\lambda\rightarrow\infty$, see
Fig.~\ref{ERAVE} (lower panel). At finite values of $\lambda$ clear
deviations are seen. In order to describe the data, we modified the
HF formula taking into account that the elastic enhancement factor
varies with $\lambda$,
\begin{equation}
\langle \sigma_{{\rm fl}}^{ab} \rangle=\left[ 1+\delta^{ab} (F-1)
\right] \frac{T}{M} \left( 1-\frac{1}{M}\right).   \label{49}
\end{equation}
As one can see, this expression gives a satisfactory description of
the data, with the numerically computed values of $F$. The problem
of an analytical dependence of $F$ on the interaction strength
$\lambda$ remains open. To shed light on this problem, we performed
a specific study of the elastic cross section in dependence on
$\lambda$ for fixed value $\kappa =0.8$ in the overlapping regime,
see Fig.~2. As one see, there is a sharp decrease of the cross
section in the transition from regular to chaotic intrinsic motion,
$\lambda \approx \lambda_{{\rm cr}}$. This result is quite
instructive since it shows how the scattering properties are
influenced by the onset of chaos in an internal dynamics. The
non-trivial point is that the analytical estimate of $\lambda_{{\rm
cr}}$ was obtained for a closed system, $\kappa =0$. However, even
in the regime of a strong coupling to the continuum, $\kappa = 0.8$,
this estimate gives a correct value for the interaction strength at
which a drastic change of scattering properties occurs.

\vspace{-0.5cm}
\begin{figure}[h!]
\includegraphics[width=7cm,angle=-90]{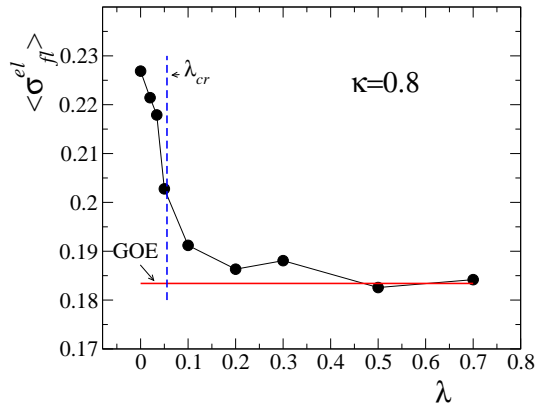}
\vspace{-0.5cm} \caption{(Color online) Fluctuational elastic cross
section as a function of the interaction strength $\lambda$ for
$M=10$ and $\kappa=0.8$ (connected circles). The horizontal line
refers to the GOE value, and dashed vertical line shows the critical
value $\lambda_{{\rm cr}}$ for the transition to chaos in the TBRE,
see Eq. (\ref{estimate}).} \label{ELv0}
\end{figure}

\section{Fluctuations of widths and resonance amplitudes}

Here we discuss the conventional assumption that for a large number
of channels the deviations of the widths from their average are
small, ${\rm Var}(\Gamma) /\langle\Gamma\rangle^2 \ll 1$, and,
therefore, for analytical estimates one can set $\Gamma_r
\approx\langle\Gamma\rangle$, see Eq. (\ref{26}). It is usually said
in justification of this assumption \cite{ericson63} that in the
overlapping regime the width can be presented as a sum of partial
widths, $\Gamma_r= \sum_{c=1}^M \Gamma_{r}^{c}$. Assuming that
individual partial widths obey the Potter-Thomas distribution, the
total width is expected to have a $\chi^2_M$ distribution, so that
${\rm Var} (\Gamma)/\langle\Gamma\rangle^2 =2/M$ is small for $M\gg
1$. In fact, it is sufficient to accept that the partial widths are
independent random variables; then ${\rm Var}(\Gamma) \propto M$ and
$\langle\Gamma\rangle\propto M$, so that ${\rm Var}(\Gamma)/\langle
\Gamma \rangle^2 \propto M^{-1}$.

However, in our previous work \cite{celardo1} we have showed that
for large values of $\kappa$ the distribution of the widths
strongly differs from the $\chi^2_M$ distribution. Our new data in
Fig.~3 give more details concerning this problem. These data were
obtained for a large number $N_r=100$ realizations of the
Hamiltonian matrices, in order to have reliable results.

The data show that as $\kappa$ increases the normalized variance,
${\rm Var}(\Gamma)/\langle\Gamma\rangle^2$, also increases,
remaining very large even for $M=20$. Moreover, the deviations
from the expected $1/M$ behavior are clearly seen signaling the
presence of correlations in the partial widths. From Fig.~3 one
can also understand how the value of ${\rm Var}(\Gamma)/\langle
\Gamma\rangle^2$ depends on the degree of intrinsic chaos
determined by the parameter $\lambda$. Specifically, for small
$\kappa$ there is no dependence on $\lambda$ and ${\rm
Var}(\Gamma)/\langle\Gamma\rangle^2$ decreases as $2/M$ for all
the ensembles, as expected. However, as $\kappa$ grows, the
dependence on $\lambda$ emerges: the weaker the intrinsic chaos
(and, consequently, the more ordered is the intrinsic spectrum)
the larger are the width fluctuations.

\begin{figure}[h!]
\includegraphics[width=7.5cm,angle=-90]{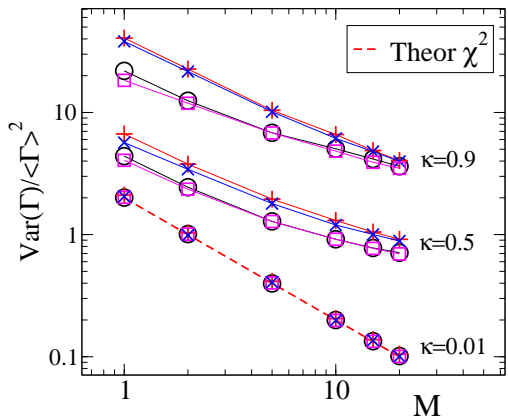}
\vspace{-0.5cm} \caption{(Color online) Normalized variance of the
width as a function of the number of channels $M$, for different
coupling strengths $\kappa$ (connected symbols are the same as in
Fig.~\ref{ERAVE}). While for small coupling, $\kappa=0.01$, the
variance decreases with the number of channels very fast in
accordance with the expected $\chi^2$-distribution (dashed line),
for large couplings, $\kappa=0.5$ and 0.9, the behavior is different
from the $1/M$-dependence.} \label{VVG}
\end{figure}

\begin{figure}[h!]
\vspace{-0.5cm}
\includegraphics[width=7cm,angle=-90]{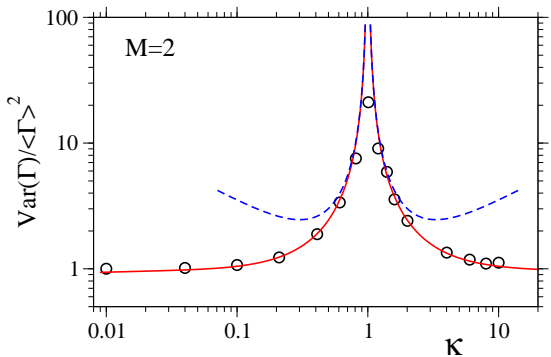}
\vspace{-0.2cm} \caption{(Color online) Numerical data for the
normalized variance of the widths vs $\kappa$ for GOE and $M=2$
(circles), in comparison with the result of numerical integration
of Eq.~(\ref{50}) (solid curve), and with Eq.~(\ref{51}) (dashed
curve), see in the text. } \label{M2int}
\end{figure}

In Ref. \cite{SFT99} the widths distribution for the GOE ensemble
was found analytically for any number of channels in the limit of
$N\to\infty$ and $M$ fixed. The general analytical result is given
in terms of a complicated threefold integral. A simpler expression
is obtained for a specific case $M=2$, for which the distribution
of the widths is expressed as a double integral,
\begin{equation}
\langle y^2\rangle=C\int_{1}^{g}\frac{d \nu}{\sqrt{\nu^2-1}}
\int_{-1}^1  d \mu \frac{(1-\mu^2)}{(\nu+g-2\mu)(\nu-\mu)^2},
                                                      \label{50}
\end{equation}
where $C=1/(2\sqrt{g^2-1}) $, $y= \pi \Gamma/D$, $g=(2/T)-1$ and $T$
is the transmission coefficient, so that from Eq. (\ref{20}) we have
$g=(1+\kappa^2)/2 \kappa$, or $\kappa= g \pm \sqrt{g^2-1}$. Note
that $\nu+g-2\mu>0$ and for $g =1$ we have $\kappa=1$. The result of
numerical integration in Eq.~(\ref{50}) is shown in Fig.~\ref{M2int}
by a solid curve. The comparison with the normalized variance of the
widths for the GOE case and $M=2$ (circles) shows a good agreement
except for the vicinity of $\kappa=1$. The difference at this point
is due to the finite $N$ effects.

\vspace{-0.5cm}

A specific interest is in the behavior of the variance at the
transition region, $\kappa \approx 1$. The analytical expression,
derived from Eq. (\ref{50}) for this region has the form,
\begin{equation}
\frac{{\rm Var}(\Gamma)}{\langle\Gamma\rangle^2}=\frac{2(2+\pi)}
{\sqrt{2(g+1)} (g-1)} \left( \ln\frac{g-1}{{g+1}} \right)^2-1,
                                                \label{51}
\end{equation}
and is shown in Fig.~4 by the dashed curve. From the above
relation for $M=2$ one can obtain that the normalized variance
diverges as
\begin{equation}
\frac{{\rm Var}(\Gamma)}{\langle\Gamma\rangle^2} \propto
\frac{1}{(1-\kappa)^2} \left[\ln(1-\kappa)\right]^2. \label{52}
\end{equation}

Our numerical simulations confirm that the divergence remains for
any number of channels and for any value of $\lambda$. Thus,
contrary to the traditional belief, the variance of widths does
not become small for a large number of channels. Another result is
that the assumption of the absence of correlations between the
resonance amplitudes $\delta z_r$ and the widths $\Gamma_r$ seems
to be incorrect in the region of a strong resonance overlap.
Indeed, the data reported in Fig.~5 demonstrate that in contrast
with the case of weak coupling, $\kappa = 0.001$, for a strong
coupling there are systematic correlations between $\delta z_r$
and $\Gamma_r$. These correlations are increasing with an increase
of the coupling strength, $\kappa$, this effect is missed in the
conventional description.

\vspace{-0.5cm}
\begin{figure}[h!]
\vspace{-0.5cm}
\includegraphics[width=7.5cm,angle=-90]{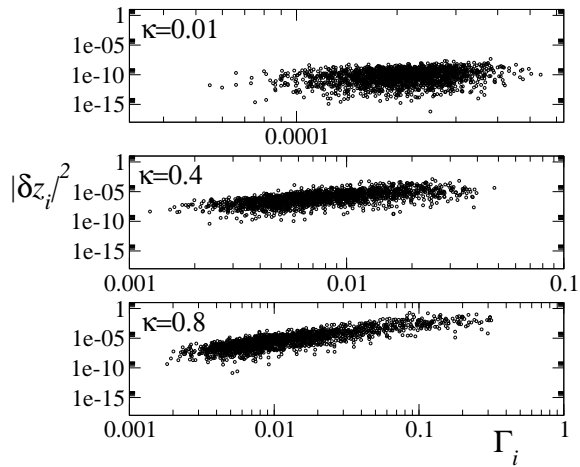}
\caption{(Color online) Absolute squares of resonance amplitudes,
$|\delta z_r|^2$, versus the widths $\Gamma_r$ for the GOE with
$M=20$. As $\kappa$ increases, the correlations between $|\delta
z_r|^2$ and $\Gamma_r$ grow.}
                                                    \label{RA}
\end{figure}

\section{Statistics of cross sections}

\subsection{Distribution of fluctuational cross sections}

According to the standard Ericson theory, the fluctuating
scattering amplitude can be written as ${\cal T}_{{\rm
fl}}^{ab}=\eta+i\xi$, where $\eta$ and $\xi$ are Gaussian random
variables with zero mean and equal variances. Since the
fluctuating cross section is given by
\begin{equation}
\sigma_{{\rm fl}}=|{\cal T}_{{\rm fl}}|^2=|\eta|^2+|\xi|^2,
                                                 \label{53}
\end{equation}
then $\sigma_{{\rm fl}}$ should have a $\chi^2$ distribution with
two degrees of freedom, that is an exponential distribution,
\begin{equation}
P(x)= e^{-x}, \quad x=\frac{\sigma_{{\rm fl}}}{\langle\sigma_{{\rm
fl}}\rangle}.                                 \label{54}
\end{equation}
This should be  valid both for the elastic and inelastic cross
sections.

\begin{figure}[h!]
\includegraphics[width=7.5cm,angle=-90]{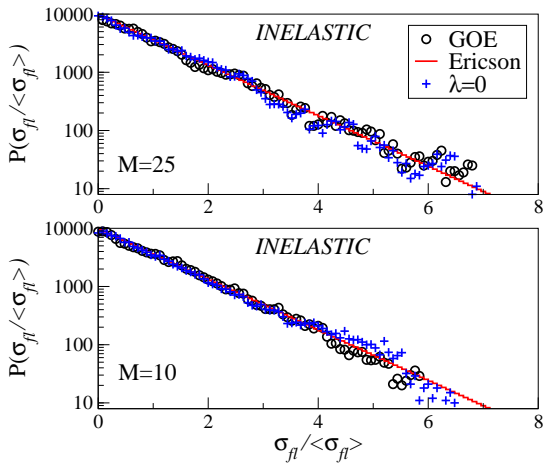}
\caption{(Color online) Distribution of the inelastic
fluctuational cross section for the GOE and for the $\lambda=0$
case, for $M=10;\,25$ number of channels and fixed $\kappa=0.9$. }
\label{ine}
\end{figure}

\begin{figure}[h!]
\includegraphics[width=7.5cm,angle=-90]{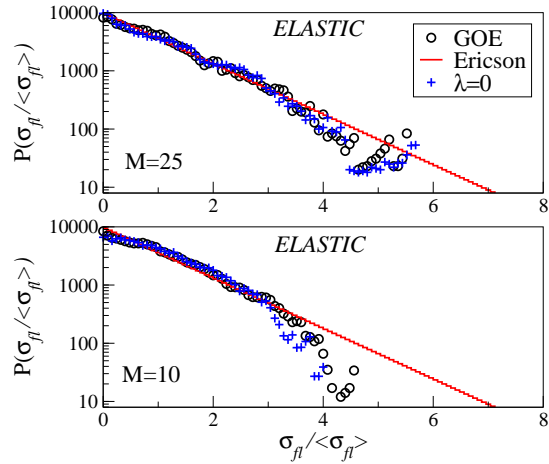}
\caption{(Color online) The same as Fig.6, but for the elastic
fluctuational cross section. A clear difference from the exponential
distribution is seen in the tails.} \label{el}
\end{figure}

In Figs. \ref{ine} and \ref{el} we show the distribution of
fluctuational cross sections for the elastic and inelastic cross
sections, with two different numbers of channels, $M=10$ and $M=25$.
Analyzing these data, one can draw the following conclusions. First,
for the inelastic cross section the data seem to follow the
predicted exponential distribution. It should be noted, however,
that a more detailed analysis with the help of the $\chi^2$-test
reveals the presence of strong deviations.

The situation with the fluctuational elastic cross section is
different due to strong deviations from the exponential distribution
occurring even for a quite large $M=25$. The fact that large
deviations from the conventional theory (for finite values of $M$)
should be expected in the elastic case were recognized also in
Refs.~\cite{DB1,DB2}. The comparison between $M=10$ and $M=25$ cases
indicate that it is natural to assume that with a further increase
of $M$ both the distributions will converge to the exponential one.
It is important to note that there is a weak dependence on the
interaction strength $\lambda$ between the particles. This is
confirmed by a closer inspection of the data of Fig.~\ref{el}.
Specifically, the data clearly show that there is a systematic
difference for the two limiting cases of zero and infinitely large
values of $\lambda$. Our results for the normalized variance (see
below), indeed, confirm a presence of this weak dependence on
$\lambda$.

The above data for the distribution of the normalized fluctuational
cross section may be treated as a kind of confirmation of the
Ericson fluctuation theory. However, it should be stressed that if
we are interested in the fluctuations of non-normalized cross
sections (at least, for elastic cross sections), one should take
into account the dependence on $\lambda$. Currently no theory allows
one to obtain the corresponding analytical results, even for the
situation where the number of channels is sufficiently large.

\subsection{Fluctuations}

Here we compare our results for the variance of cross sections with
the Ericson fluctuations theory \cite{ericson63}, and with more
recent results for the GOE \cite{DB1,DB2}. According to the standard
predictions, the variance of fluctuations of both elastic and
inelastic cross sections,
\begin{equation}
{\rm Var}(\sigma^{ab})=\langle(\sigma^{ab}-
\langle\sigma^{ab}\rangle)^2\rangle,              \label{55}
\end{equation}
is directly connected to the average cross sections by
Eq.~(\ref{31}). It is useful to express the variance of the cross
sections in terms of the scattering matrix. In our statistical model
for $a \ne b$, $\langle{\cal T}^{ab}\rangle=i\langle
S^{ab}\rangle=0$. Therefore, the variance of the inelastic cross
section reads
\begin{equation}
{\rm Var}(\sigma^{ab})=\langle\sigma_{{\rm fl}}^{ab}\rangle^2
=\langle|S^{ab}_{{\rm fl}}|^2\rangle^2. \label{56}
\end{equation}

For the elastic scattering one can write $\langle {\cal T}^{aa}
\rangle=-i (1-\langle S^{aa}\rangle)$, so that $\sigma_{{\rm
dir}}=|1-\langle S^{aa} \rangle|^2$, and $\sigma_{{\rm fl}}=\langle
|S^{aa}_{{\rm fl}}|^2 \rangle$. Therefore, for the variance of the
elastic cross sections, one obtains,
$$
{\rm Var}(\sigma^{aa})=2 \langle\sigma_{{\rm fl}}^{aa}\rangle
\langle \sigma_{{\rm dir}}^{aa}\rangle+\langle\sigma_{{\rm
fl}}^{aa}\rangle^2 =
$$
\begin{equation}
=\langle|S^{aa}_{{\rm fl}}|^2\rangle^2+2|1-\langle
S^{aa}\rangle|^2 \langle |S^{aa}_{{\rm fl}}|^2 \rangle, \label{57}
\end{equation}

\begin{figure}[h!]
\includegraphics[width=7.0cm,angle=-90]{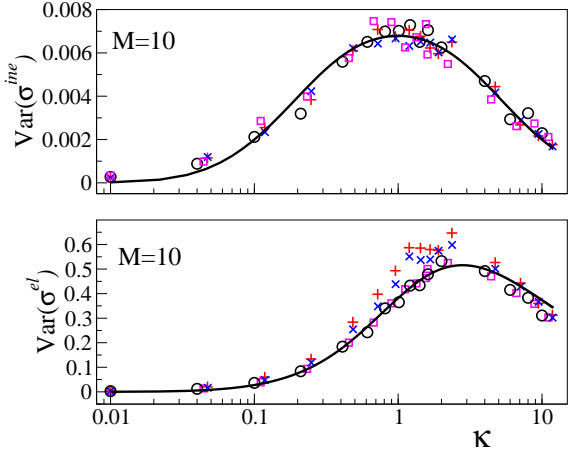}
\caption{(Color online) Variance of inelastic (upper panel) and
elastic (lower panel) cross sections, see Eq.~(\ref{25}), for $M=10$
as a function of $\kappa$ for different interaction strengths: the
GOE (circles), $\lambda=0$ (pluses), $\lambda\rightarrow\infty$
(squares), and $\lambda=1/30$ (crosses). For the comparison the
theoretical curve for $MT \gg 1$, obtained for the GOE from
Eqs.~(\ref{60},\ref{61}) (solid curve) is also added. Note that the
maximum of ${\rm Var}(\sigma^{el})$ is shifted from $\kappa=1$ due
to the presence of direct processes.} \label{fluctELINE}
\end{figure}

The discussed above conventional predictions and our analysis of the
average cross section in Sec. VI imply that the variances of cross
sections depend on the intrinsic interaction strength $\lambda$
through the average cross sections. Our data for a relatively large
number $M=10$ of channels, see Fig.~\ref{fluctELINE}, indeed,
correspond to this expectation. Since for the inelastic scattering
the average cross section does not depend on the interaction
strength, it is quite expected that the same occurs for the variance
of the inelastic cross section. Our data confirm this expectation.
On the other hand, the variance of the elastic cross section reveals
a clear dependence on the value of $\lambda$. As one can see, this
dependence is quite strong in the region of strongly overlapped
resonances, for $\kappa \approx 1$.

Let us now compare our data with the exact expressions for the
variance of cross sections. To do this, it is convenient to express
this variance in terms of the scattering matrix,
$$
{\rm Var}(\sigma^{ab})=\langle|S^{ab}_{{\rm fl}}|^4\rangle-
\langle|S^{ab}_{{\rm fl}}|^2\rangle^2+
$$
\begin{equation}
-\delta^{ab}\Bigl(2\Bigl[(1-\langle S^{aa\ast}\rangle)
\langle|S^{aa}_{{\rm fl}}|^2 S^{aa}_{{\rm fl}}\rangle + {\rm c.c.}
\Bigr]-2 |1-\langle S^{aa}\rangle|^2 \langle|S^{ab}_{{\rm fl}}|^2
\rangle\Bigr).                                      \label{58}
\end{equation}
Comparing Eq. (\ref{58}) with the standard predictions, Eqs.
(\ref{56},\ref{57}), one can see that they are correct if:
$$
({\rm i})\;\langle|S^{ab}_{{\rm fl}}|^4\rangle
-2\langle|S^{ab}_{{\rm fl}}|^2\rangle^2=0,
$$
\begin{equation}
({\rm ii})
\;\langle S_{{\rm fl}}^{aa}|S^{ab}_{{\rm fl}}|^2\rangle=0.
                                                 \label{59}
\end{equation}
These properties are consistent with the Gaussian character of the
distribution for the fluctuational scattering matrix.

The analytical expressions for the variance of elastic and inelastic
cross sections were obtained in Refs. \cite{DB1,DB2} for the GOE
case, any number of channels and any coupling strength with the
continuum. However, simple expressions were derived only for $MT \gg
1$. Even under such a condition, the analytical results show
deviations from the conventional assumptions. Specifically, it was
found,
$$
({\rm i})\; \langle|S^{ab}_{{\rm fl}}|^4\rangle-2\langle|S^{ab}_{{\rm
fl}}|^2\rangle ^2=
$$
\begin{equation}
=(1+7\delta^{ab})[6-4(T^a+T^b)+r_2] \,\frac{2(T^aT^b)^2}{(S_1+1)^3},
                                                 \label{60}
\end{equation}
and
\begin{equation}
({\rm ii})\; \langle|S^{aa}_{{\rm fl}}|^2 S^{aa}_{{\rm fl}}\rangle= -8\langle
S^{aa\ast}\rangle\,\frac{(T^a)^3}{(S_{1}+1)^2}, \label{61}
\end{equation}
where $S_1=\sum T^c$, $S_2=\sum (T^c)^2$ and $r_2=(S_2+1)/(S_1+1)$.

The theoretical values for the variance of the cross sections
obtained from Eqs.(\ref{60},\ref{61}) and from the HF formula,
through Eq.(\ref{58}), are shown in Fig.~\ref{fluctELINE} by solid
curve. The agreement is good for the GOE case in the strong coupling
regime, as expected. As we can see from Eqs. (\ref{60},\ref{61}),
assumptions of Eq. (\ref{59}) are valid for large $S_1$. In
particular it was shown in \cite{DB1,DB2} that the ratio ${\rm
Var}(\sigma_{{\rm fl}})/\langle\sigma_{{\rm fl}}\rangle^2$, being
equal to one in standard theory, significantly differs from unity in
the range $10<MT<20$, where this theory is expected to be valid.

\begin{figure}[h!]
\vspace{-0.5cm}
\includegraphics[width=7.5cm,angle=-90]{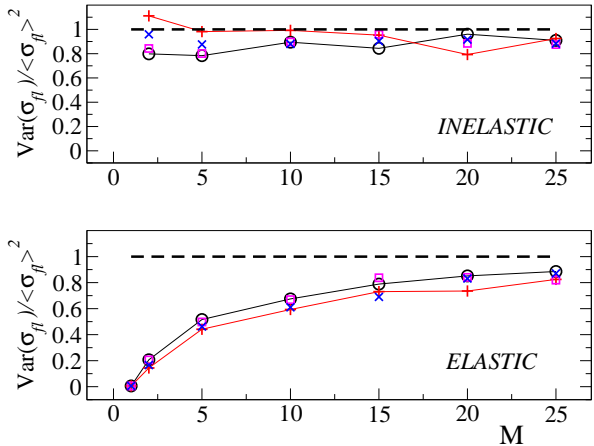}
\vspace{-0.5cm} \caption{(Color online) Normalized variance for
$\kappa=0.9$ versus the number of channels for the elastic and
inelastic cross sections: GOE case (connected circles),
$\lambda=0$ (connected pluses), $\lambda\rightarrow\infty$
(squares), and $\lambda=1/30$ (crosses). The theoretical value in
the Ericson theory, ${\rm Var}(\sigma)/\langle\sigma\rangle^2=1$,
is shown as a dashed line. } \label{Vars}
\end{figure}

It is now instructive to see how the normalized variance of the
cross sections depends on the number of channels, see Fig.
\ref{Vars}. According to the Ericson prediction, the ratio of the
variance to the square of the mean of the cross sections has to be 1
for strongly overlapped resonances, in the limit of large number of
channels. As one can see, the data for the inelastic scattering
roughly confirm this prediction. It is not a surprise that
practically there is no dependence on the strength $\lambda$ of
interaction between the particles. On the other hand, there is a
small systematic difference from the predicted value, that emerges
for all values of $\lambda$, as well as for the GOE case. One can
expect that this difference disappears for a much larger number of
channels.

A more interesting result arises for the elastic cross section. As
one can see from Fig. \ref{Vars}, in this case there is a strong
difference from the limiting value of one even in the case when the
number of channels is relatively large, $M=10-20$. The data clearly
indicate that for the applicability of the standard theory one needs
to have a very large number of channels, at least larger than
$M=25$.

Another instructive observation is a weak dependence of the
normalized variance on the interaction strength $\lambda$. This
result is in agreement with the data reported in Fig. \ref{el} for
the distribution of individual values of the cross section, where a
systematic deviation can be seen when comparing the GOE case with
the case of $\lambda=0$. We would like to stress that the weak
$\lambda$-dependence is in contrast with a strong dependence
occurring for the non-normalized variance, see Fig.
\ref{fluctELINE}. One can treat this effect as manifesting that both
the variance and the square of the cross section average depend on
$\lambda$ practically in the same way. Therefore, their ratio turns
out to be almost independent on $\lambda$. As one can see, although
the standard predictions are not correct for the non-normalized
variance, they are in a good correspondence with the data for the
normalized variance.

\subsection{Correlation functions}

Here we compare our results for the correlation function of cross
sections and the scattering matrix with the standard predictions,
see Section V. The correlation functions for the cross section and
for the scattering matrix were computed according to Eqs.
(\ref{32},\ref{34}) for the elastic and inelastic cross sections.
The correlation lengths for the cross section, $l_{\sigma}$, and
for the scattering matrix, $l_{S}$, are defined as the energy for
which the correlation function is $1/2$ of its initial value. Our
results can be summarized as follows.

(A) For large $M$, we found $l_{\sigma} \approx l_{S}$ for any
interaction strength $\lambda$. On the contrary, for smaller $M$,
our data show that $l_{S}<l_{\sigma}$, and this difference grows for
the weaker interaction between the particles, $\lambda$.

\begin{figure}[h!]
\includegraphics[width=7.5cm,angle=-90]{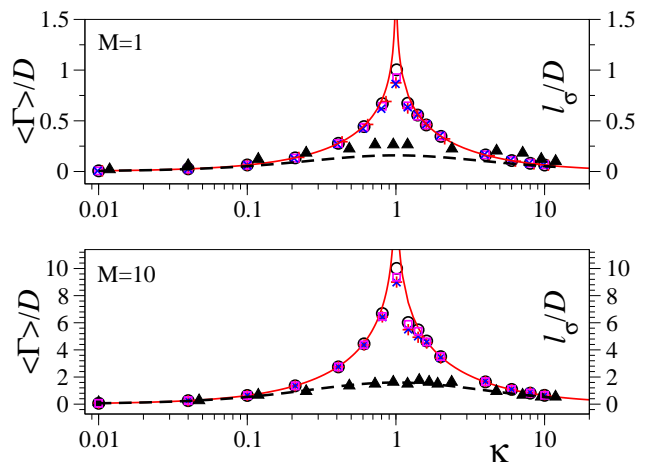}
\vspace{-0.5cm} \caption{(Color online) Average width, $\left<
\Gamma \right>$, and correlation length, $l_\sigma$, normalized to
the mean level spacing at the center of the spectra, versus $\kappa$
for $M=1$ (upper panel) and $M=10$ (lower panel). Solid curves show
the MS-expression (\ref{61}). Open circles refer to the GOE case,
pluses to $\lambda=0$, squares to $\lambda \rightarrow\infty$, and
crosses to $\lambda=1/30$, all for the normalized average width.
Full triangles stand for the normalized correlation length at
$\lambda=1/30$. The dashed line shows the Weisskopf relation
(\ref{60}).} \label{924GD10}
\end{figure}

(B) For large $M$, the correlation functions are Lorentzian for all
$\lambda$, while for a small number of channels the correlation
function is not Lorentzian, in agreement with the results of
\cite{dittes92}. Moreover, for any $M$, the correlation length is
different from the average width, as one can expect due to Eq.
(\ref{37}), apart from the region of small $\kappa$, see Fig.
\ref{924GD10}. For a large number of channels, the correlation
length is determined instead by the transmission coefficient through
the Weisskopf relation, see \cite{lehmann95} and references therein,
\begin{equation}
\frac{l}{D}=\frac{MT}{2 \pi}=\frac{M}{2 \pi} \frac{4 \kappa}{(1+\kappa)^2}.
                                                               \label{62}
\end{equation}

\vspace{-1.0cm}

\vspace{-0.5cm}
\begin{figure}[h!]
\includegraphics[width=8cm,angle=-90]{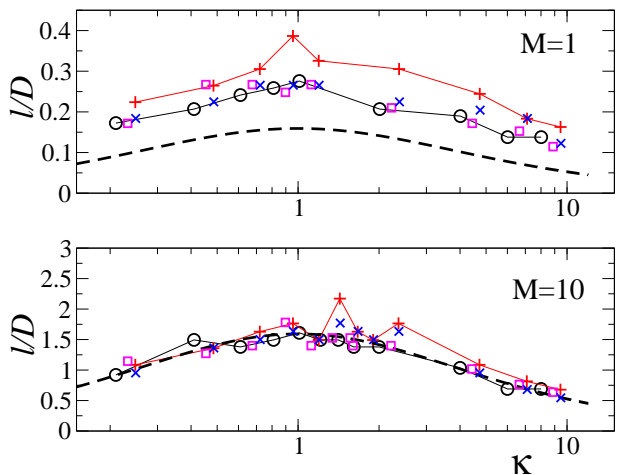}
\vspace{-1.0cm} \caption{(Color online) Elastic-elastic
correlation length of the cross section at different values of
$\lambda$ as a function of $\kappa$ for $M=1$ and $M=10$; the GOE
case (connected circles), $\lambda=0$ (pluses connected by a
line), $\lambda\rightarrow\infty$ (squares), and $\lambda=1/30$
(crosses). The Weisskopf relation (\ref{62}) is shown by dashed
curve. } \label{Ccorr}
\end{figure}

In  Fig. \ref{Ccorr}, it is shown that, for a large number of
channels, the elastic correlation length is in agreement with
Eq.(\ref{62}) for all values of the interaction strength, $\lambda$.
The same occurs for the inelastic correlation length.

The Weisskopf relation (\ref{62}) has been also derived in Ref.
\cite{agassi75} for small values of the ratio $m=M/N$, in the
overlapping regime for the TBRE ensemble with the infinite
interaction, as well as in Ref. \cite{verbaarschot85} for the GOE
ensemble. In \cite{lehmann95} the correlation function for the GOE
ensemble was computed also when $m=M/N$ is not small, and the
deviations from Eq. (\ref{62}) and from the Lorentzian form of the
correlation function were found. This is not in contrast with our
results for the case of small ratio of $m$, see discussion in Sec.
IV.

The fact that the correlation length is not equal to the average
width was recognized long ago, see Refs. \cite{moldauer75} and
\cite{brody81}. However, the statements based on the equality
(\ref{37}) still appear in the literature, see, for example, Ref.
\cite{abfalterer00}. The relation between the average resonance
width and the transmission coefficient is given by the
Moldauer-Simonius (MS) formula \cite{moldauer67,simonius74},
\begin{equation}
M \ln(1-T)=-2 \pi\,\frac{\langle\Gamma\rangle}{D}. \label{63}
\end{equation}
It can be seen from this expression and Eq. (\ref{62}) that the
equality $l=\langle\Gamma\rangle$ is true only for small $T$.

\section{Conclusions}

In conclusion, we have studied the statistics of cross sections
for a fermion system coupled to open decay channels. For the first
time we carefully followed various signatures of the crossover
from isolated to overlapping resonances in dependence on the
strength of inter-particle interaction modelled here by the
two-body random ensemble. The study was performed for the simplest
Gaussian ensemble of decay amplitudes. Even in this limiting case,
when these amplitudes were considered as uncorrelated with the
intrinsic dynamics, we found significant dependence of reaction
observables on the strength of intrinsic interaction. We expect
this dependence to be amplified with realistic interplay of decay
amplitudes and internal wave functions. Such studies should be
performed in the future.

A detailed comparison has been carried out of our results with
standard predictions of statistical reaction theory. The average
cross section was compared with the Hauser-Feshbach formula for a
large number of channels. In the inelastic case this description
works quite well in the overlapping resonance regime for any
interaction strength, while in the elastic case strong deviations
have been found if the intrinsic motion is not fully chaotic.

The study of Ericson fluctuation theory shows that the assumption
that the fluctuations of the resonance widths become negligible
for a large number of channels is wrong in the overlapping regime.
We found that the fluctuations of resonance widths increase with
the coupling to the continuum, and we gave evidence that the
relative fluctuation of the width (the ratio of the variance to
the square of the average width) diverges at $\kappa=1$ for any
number of channels. This should imply that for any number of
channels the differences from the standard theory should increase
as $\kappa$ increases.

In order to study the relationship between the variance of the
cross section and its average value, it is necessary to take into
account the dependence of the average cross section on the
intrinsic interaction strength $\lambda$. Even when this is done,
the standard prediction about the variance of the cross section
was found to be a good approximation only for a very large number
of channels. For $M$ between $10$ and $20$, where the Ericson
prediction could be expected to be valid, consistent deviations
have been demonstrated.
In particular, the distribution of cross sections shows that the
probability of a large value of the cross section, mainly for the
elastic case (or in the presence of direct reactions), can be well
below conventional predictions.

Finally, we have shown that, in agreement with previous studies, the
correlation length differs from the average width for any number of
channels. On the other hand, the Weisskopf relation (\ref{62}) that
connects the correlation length of the cross section to the
transmission coefficient, works, for a large number of channels, at
any value of the intrinsic interaction strength $\lambda$. In many
situations we have seen that increase of $\lambda$ in fact
suppresses the fluctuations in the continuum. This can be understood
qualitatively as a manifestation of many-body chaos that makes all
internal states uniformly mixed.

Our results can be applied to any many-fermion system coupled to the
continuum of open decay channels. The natural applications first of
all should cover neutron resonances in nuclei, where rich
statistical material was accumulated but the transitional region
from isolated to overlapped resonances was not studied in detail.
The interesting applications of a similar approach to molecular
electronics and electron tunneling spectroscopy can be found in the
recent literature \cite{cacelli07,walczak07}. Other open mesoscopic
systems, for example, quantum dots and quantum wires, should be
analyzed as well in the crossover region. One can expect very
promising designated studies for checking the statistical properties
of resonances in such controllable experiments as those in microwave
cavities and in acoustical chaos. Open boson systems in atomic traps
also can be an interesting object of future theoretical and
experimental studies.

\section{Acknowledgment}

We acknowledge useful discussion with D. Savin, T.~Kawano,
V.~Sokolov and T. Gorin. The work was supported by the NSF grants
PHY-0244453 and PHY-0555366. The work by G.P.B. was carried out
under the auspices of the National Nuclear Security Administration
of the U.S. Department of Energy at Los Alamos National Laboratory
under Contract No. DE-AC52-06NA25396. F.M.I. acknowledges partial
support by the CONACYT (M\'exico) grant No~43730.

\end{document}